\documentclass[a4paper]{article}
\usepackage{graphics,color,epsfig}

\def\linadj#1{\normalbaselines
	\multiply\lineskip#1 \divide\lineskip100
 	\multiply\baselineskip#1 \divide\baselineskip100
	\multiply\lineskiplimit#1 \divide\lineskiplimit100 }
\newcommand{\n}{\noindent}

\begin{document}

\title{\bf A Simple Statistical Model for Analysis of QGP-droplet (Fireball) Formation }
\date{}

\author{ R.Ramanathan, Y.K. Mathur, K.K. Gupta$^+$ \\ and Agam K. Jha}

\maketitle

\begin{center}

Department of Physics, University of Delhi, Delhi - 110007, INDIA

$^+$Department of Physics, Ramjas College, University of Delhi, \\ Delhi - 110007, INDIA
\end{center}

\linadj{200} 

\n{\bf  Abstract:}

We construct the density of states  for quarks and gluons using the `Thomas - Fermi model' for atoms and the `Bethe model' for nucleons as templates. With parameters to take care of the plasma (hydrodynamical) features of the QGP with a thermal potential for the interaction, we find a window in the parametric space of the model where observable QGP droplets of $ \sim $ $5$ fm radius can occur with transition temperature in the range $140$ MeV to $250$ MeV. By matching with the expectations of Lattice Gauge estimates of the QGP-hadron transitions, we can further narrow the window, thereby restricting the allowed values of the flow-parameters of the model.

\vfill
\eject

There are increasing expectations of hadron transition to a quark-gluon plasma phase at about $150-170$ MeV [1] from Lattice Calculations. There is also the natural possibility of QGP-droplet (Fireball) formation in Ultra Relativistic Heavy Ion Collision (URHIC) [2].     
 
The physics of such a QGP droplet is too complicated to be understood with a rigorous application of QCD to the problem of QGP droplet formation within a hadronic medium. This has forced several attempts at modelling the phenomenon to gain insight into the physical process of droplet formation using equation of state [2,3,4], microscopic transport equation [5,6], hydrodynamical approaches [7,8] etc. In this paper, we briefly report on a simple statistical model of QGP [11,12] which captures a good chunk of the the physics of the QGP-hadron phase transition which can be used in the phenomenological analysis of `Fireball' data as and when they are available from the URHIC experiments going on at various laboratories at present. Further, we use the thermal model potential in the construction of the density of states for the quarks and gluons in the QGP as the thermal model [9,10] has proved to be very successful in explaining the particle multiplicities measured in URHIC at the SPS.

 In this brief report we will only give an outline of our approach, reserving a detailed version to a later date. Using the Thomas-Fermi model for the atom [13] and the Bethe model for the nucleons [14] as templates, we construct the density of states of relativistic quarks and gluons as

\begin{equation}
\rho_{q,g}(k) = (v / \pi^2)\biggl[(-V_{conf}(k))^2 \biggl(\frac{dV_{conf}(k)}{dk}\biggl)\biggl]_{q,g}
\end{equation}                                                                                                                               
where $v$ is the volume occupied by the QGP and $k$ is the relativistic four momentum in natural units. $V_{conf}(k)$ could be any confining potential for quarks and gluons, but for the present we choose to work with a modified thermal potential. 

The thermal potential [10] is

\begin{equation}
[V_{\mbox{conf}}(k)]_{q,g} = (1/2k)\gamma_{q,g} ~ g^2 (k) T^2 - m_0^2 / 2k
\end{equation}                                                                                                                               
where $g^2(k)$ is the QCD coupling constant, which for quarks with three flavours is

\begin{equation}
g^2(k) = (4/3) (12\pi/27) (1/ ln(1+k^2/\Lambda^2))
\end{equation}                                                                                                                              
with the QCD parameter $\Lambda = 150~MeV$. $\gamma_{q,g}$ are the phenomenological [10] flow parameters introduced to take care of the hydrodynamical aspects of the hot QGP droplet (fireball).

 The model has a low energy cut-off 

 \begin{equation}
  k_{min} = (\gamma_{q,g} N^{1/3} T^2 \Lambda^{2}/2)^{1/4}
\end{equation}                                                                                                                               
with $$N = (4/3)(12\pi/27).$$

   With the further simplifying assumption of a pure pionic medium surrounding the QGP droplet [3], we compute the free energy of the system of non-interacting fermions (upper sign) or bosons (lower sign) at temperature T as

\begin{equation}
F_i = \mp T g_i \int dk \rho_i (k) \ln (1 \pm e^{-(\sqrt{m_{i}^2 + k^2}) /T})
\end{equation}                                                                                                                               
where $\rho_{i}(k)$ is the density of states of a particle $i$ (quarks, gluons, interface, pion etc.)being the number of states with momentum between $k$ and $k+dk$ in a spherically symmetric situation, and $g_{i}$ is the degeneracy factor (color and spin degeneracy) which is 6 for quarks, 8 for gluons and 1 for the pions and the interface.

 The interface is assumed to be a modified Weyl surface [15],

\begin{equation}
F_{surface} = \frac{1}{4}  R^{2}  T^{3} \gamma
\end{equation}                                                                                                                              
where $R$ is the radius of the droplet and $\gamma$ is a modification sought to be introduced to take care of the plasma (hydrodynamical) nature of the droplet and is consciously chosen as   

\begin{equation}
\gamma = \sqrt{2}\times \sqrt{(1/\gamma_g)^2+(1 / \gamma_q)^2},
\end{equation}                                                                                                                               
which is the inverse r.m.s value of the flow parameter of the quarks and gluons respectively.

  The pion free energy is [3]

\begin{equation}
F_{\pi} = (3 ~ T/2\pi^2 )v \int_0^{\infty} k^2 dk \ln (1 - e^{-\sqrt{m_{\pi}^2 + k^2} / T})
\end{equation}

For the quark masses we use the current (dynamic) quark masses $m_{u} = m_{d} =0$ MeV and $m_{s} = 150$ MeV

 \n{\bf Results and Conclusion : }

With all the above numerical and theoretical inputs we have computed the free energy contributions of the u + d quarks, s-quarks and for the gluons while retaining the same behaviour  for the  pions as in [3,4]. All the energy integrations involved for the quark sector have a low energy cut-off at approximately $100$ MeV (for example, it is 94.76 MeV at T = 152 MeV) by virtue of (4), and the integral saturates at an upper cut-off at nearly four times the low energy cut-off energy.

\begin{figure}
\begin{center}
\epsfig{figure=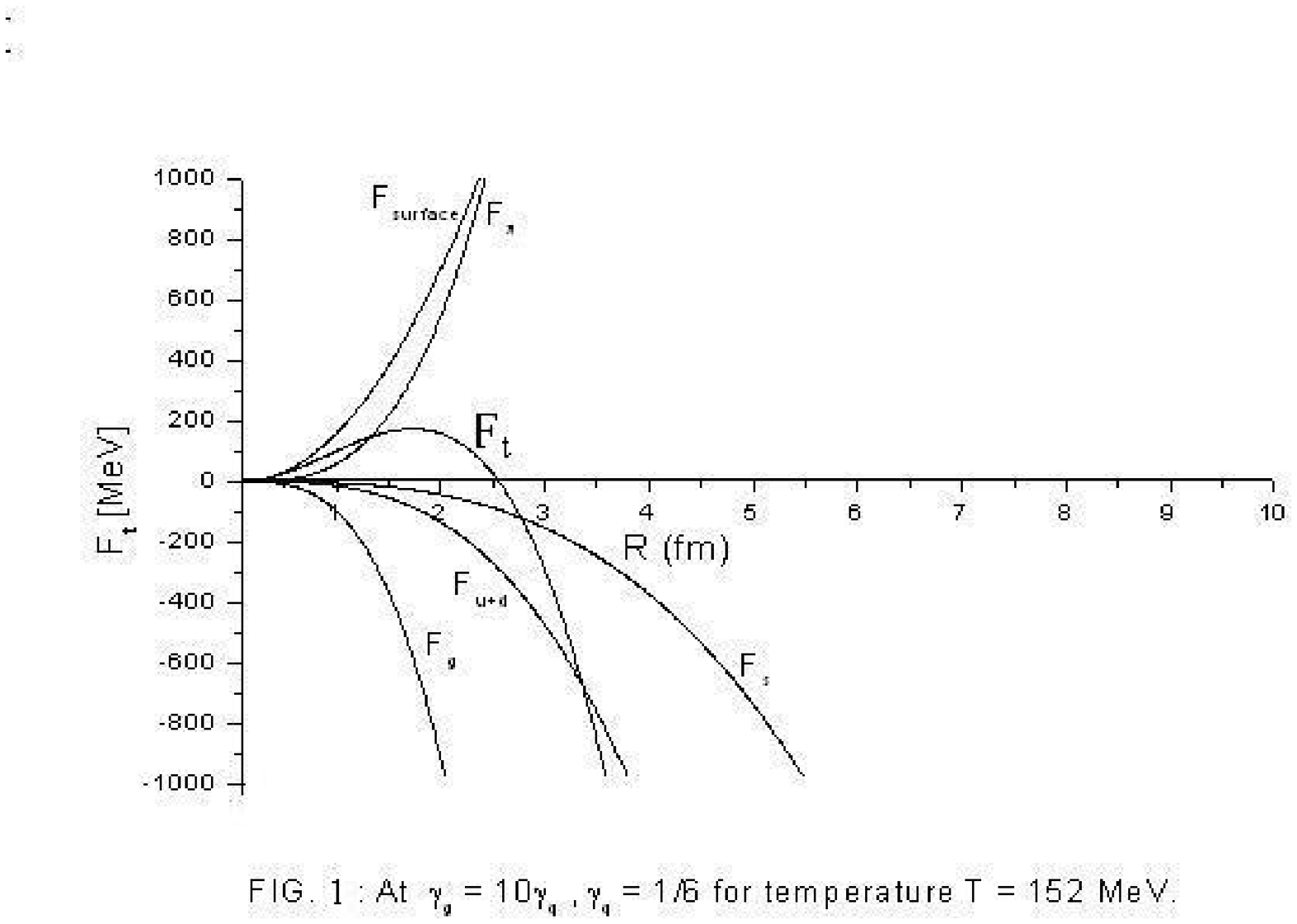,height=2.5in,width=3.5in}
\end{center}
\label{XX}
\caption{Individual contribution to Free - energy from the quarks, gluons, pions and the interface leading to the total Free - energy at $T=152$ MeV.}
\end{figure}

 \begin{figure}
 \begin{center}
 \epsfig{figure=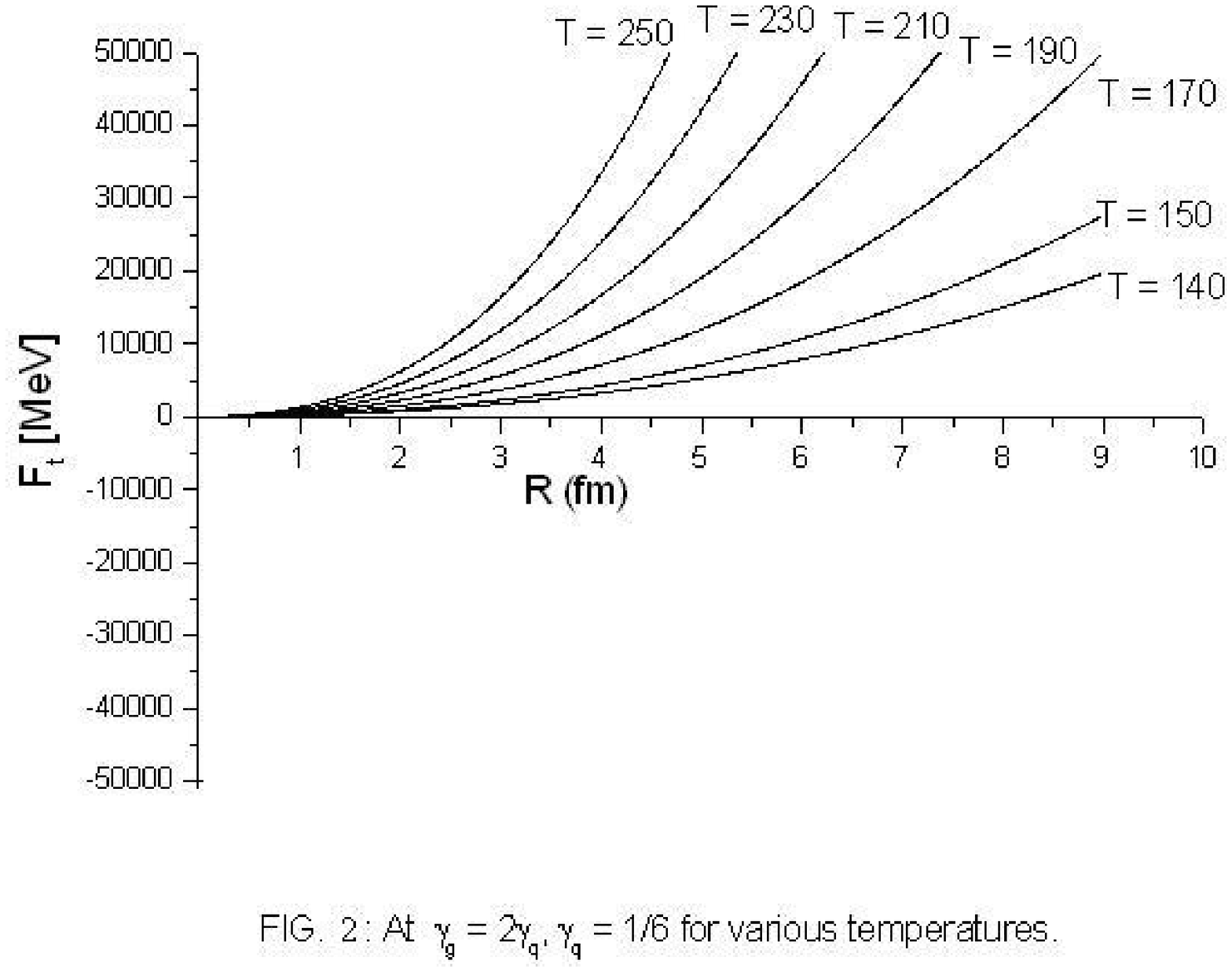,height=2.5in,width=3.5in}
\label{fig2}
\caption{ The variation of total Free-energy $F_{t}$ of the QGP droplets in a pionic medium at different temperatures for                 the flow-parameters $\gamma_{g} = 2\gamma_{q}$  and $ \gamma_q = 1/6 $ .}
\end{center}
\end{figure}

\begin{figure}
\begin{center}
\epsfig{figure=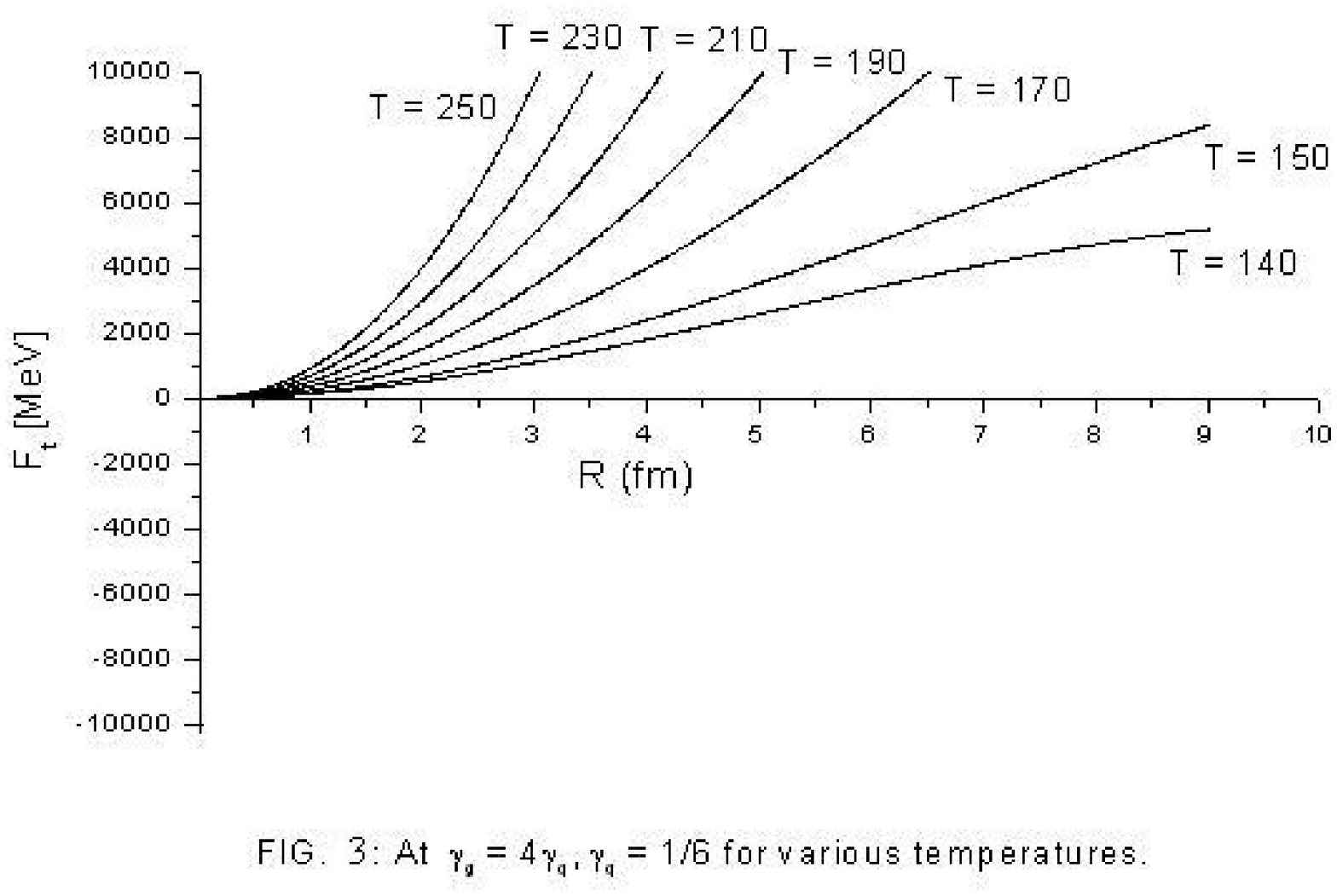,height=2.5in,width=3.5in}
\label{fig3}
\caption{The variation of total Free-energy $F_{t}$ of the QGP droplets in a pionic medium at differnt temperatures for the flow-parameters $\gamma_{g} = 4\gamma_{q} $ and $ \gamma_{q} = 1/6 $ .}
\end{center}
\end{figure}

\begin{figure}
\begin{center}
\epsfig{figure=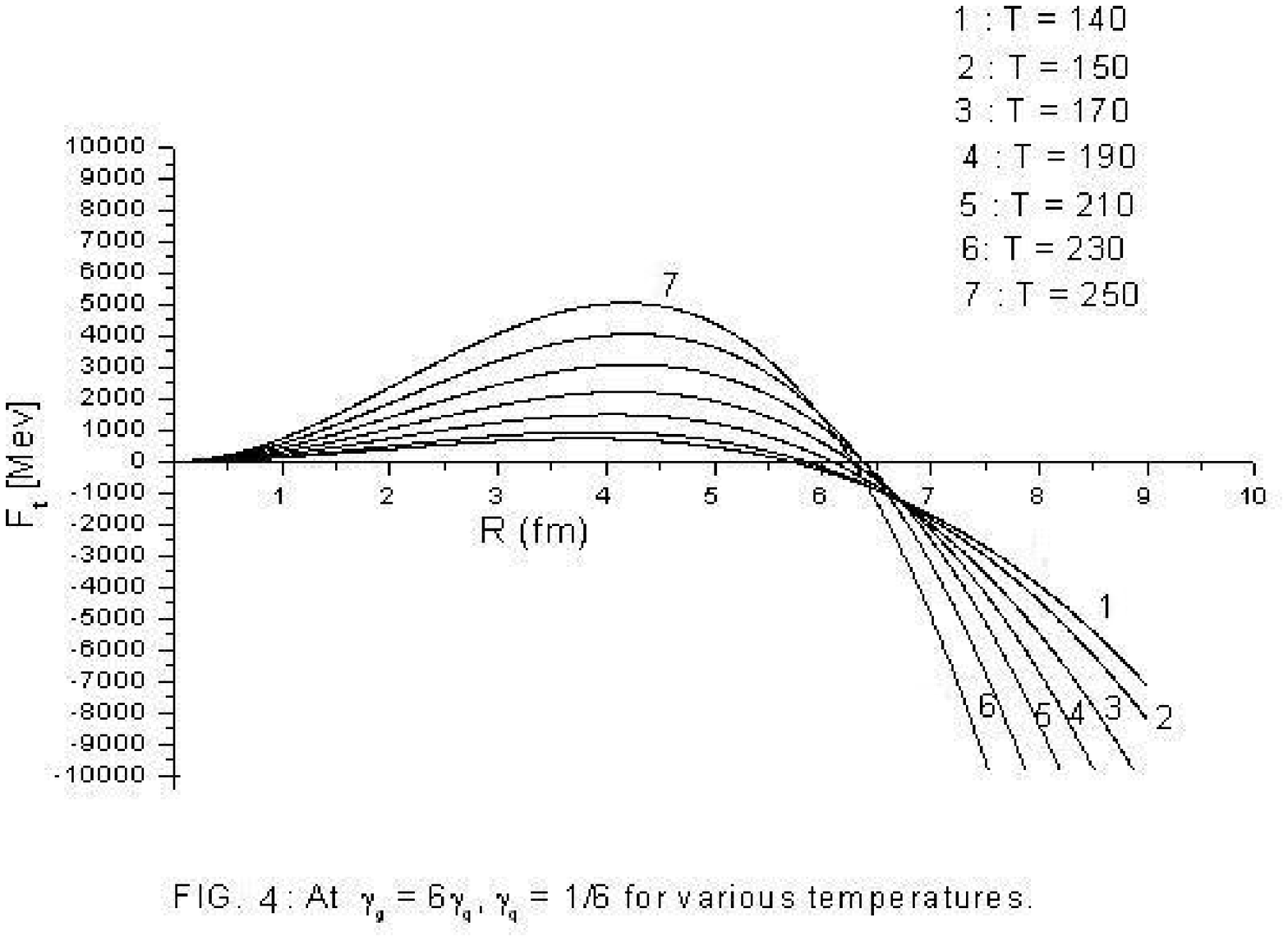,height=2.5in,width=3.5in}
\label{fig4}
\caption{The variation of total Free-energy $F_{t}$ of the QGP droplets in a pionic medium at different temperatures for the flow-parameters $\gamma_{g} = 6\gamma_{q}$ and $ \gamma_{q} = 1/6 $ .}
\end{center}
\end{figure}

\begin{figure}
\begin{center}
\epsfig{figure=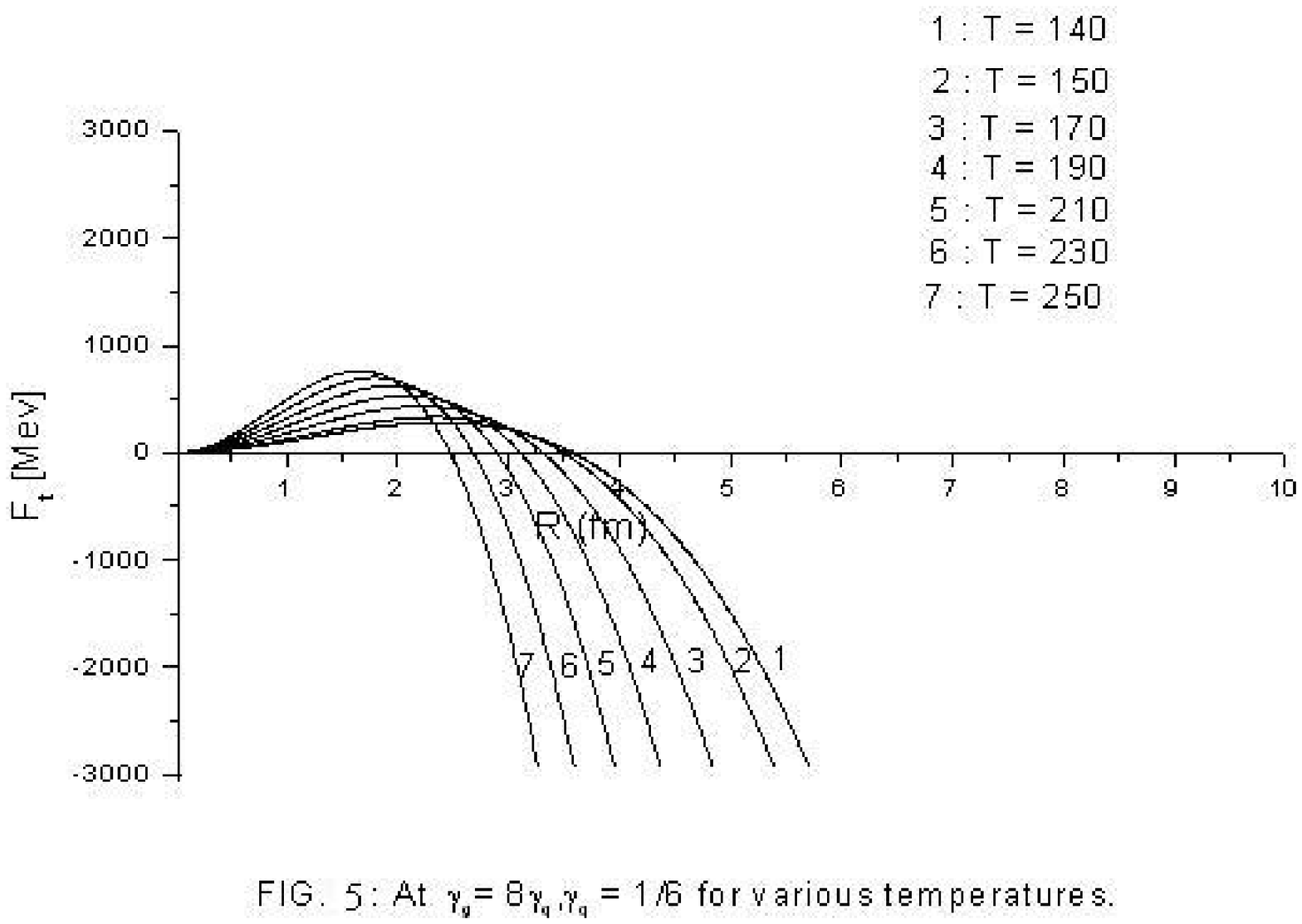,height=2.5in,width=3.5in}
\label{fig5}
\caption{The variation of total Free-energy $F_{t}$ of the QGP droplets in a pionic medium at different temperatures for the flow-parameters $\gamma_{g} = 8\gamma_{q} $ and $ \gamma_{q} = 1/6 $ .}
\end{center}
\end{figure}

\begin{figure}
\begin{center}
\epsfig{figure=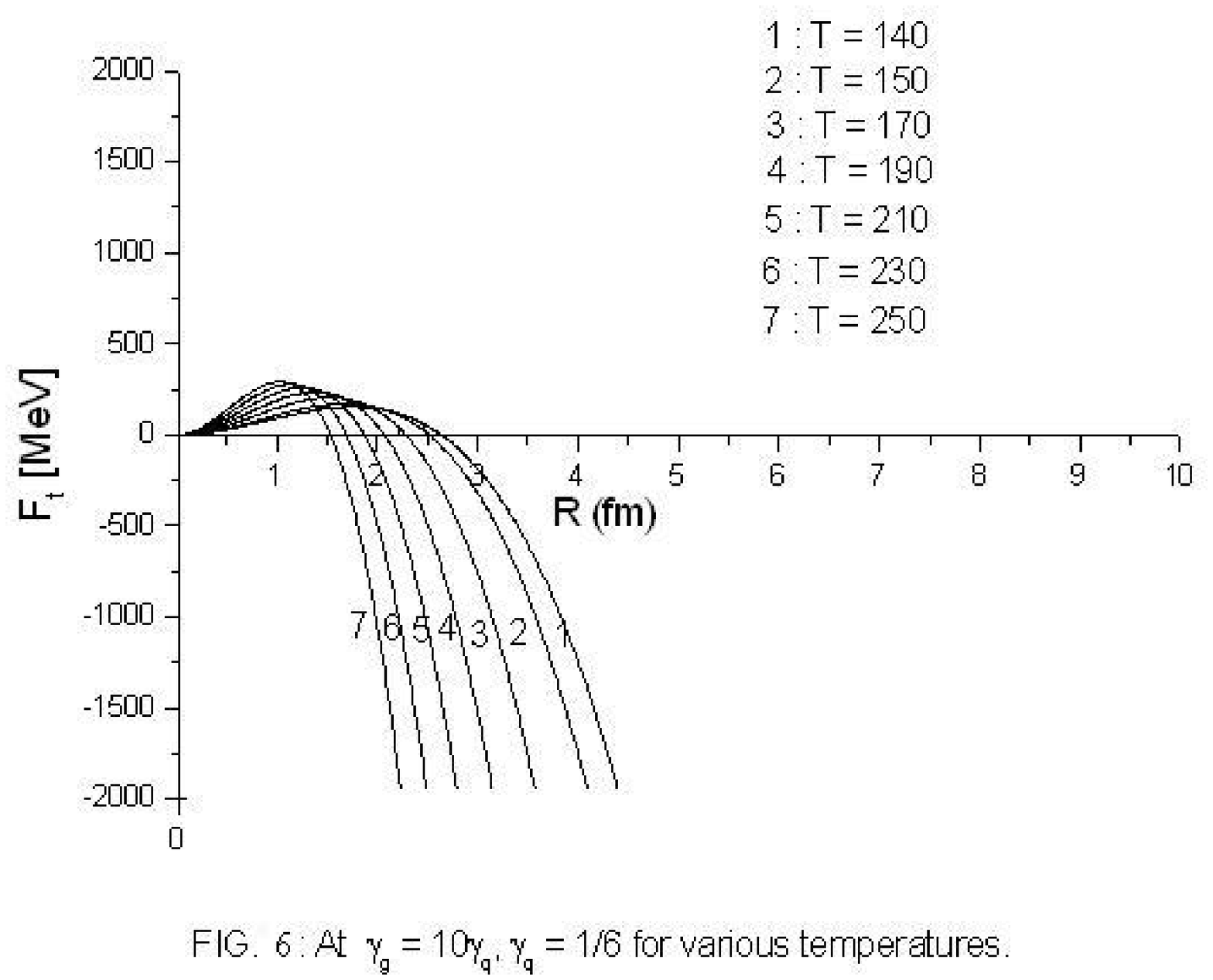,height=2.5in,width=3.5in}
\label{fig6}
\caption{The variation of total Free-energy $F_{t}$ of the QGP droplets in a pionic medium at different temperatures for the flow-parameters $\gamma_{g} = 10\gamma_{q}$ and $ \gamma_{q} = 1/6 $ .}
\end{center}
\end{figure}

\begin{figure}
\begin{center}
\epsfig{figure=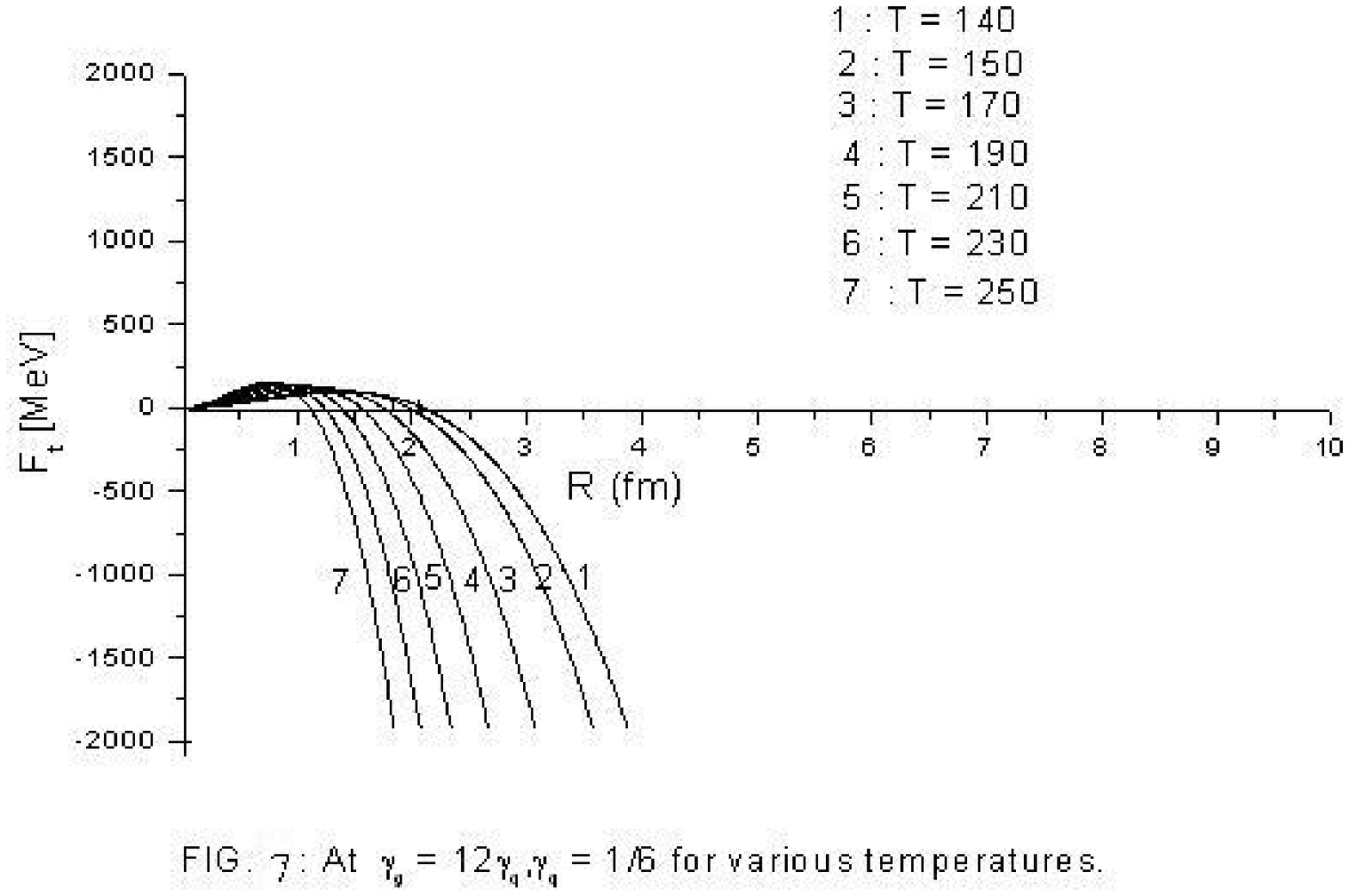,height=2.5in,width=3.5in}
\label{fig7}
\caption{The variation of total Free-energy $F_{t}$ of the QGP droplets in a pionic medium at different temperatures for the flow-parameters $\gamma_{g} = 12\gamma_{q}$ and $ \gamma_{q} = 1/6$ .}
\end{center}
\end{figure}

\begin{figure}
\begin{center}
\epsfig{figure=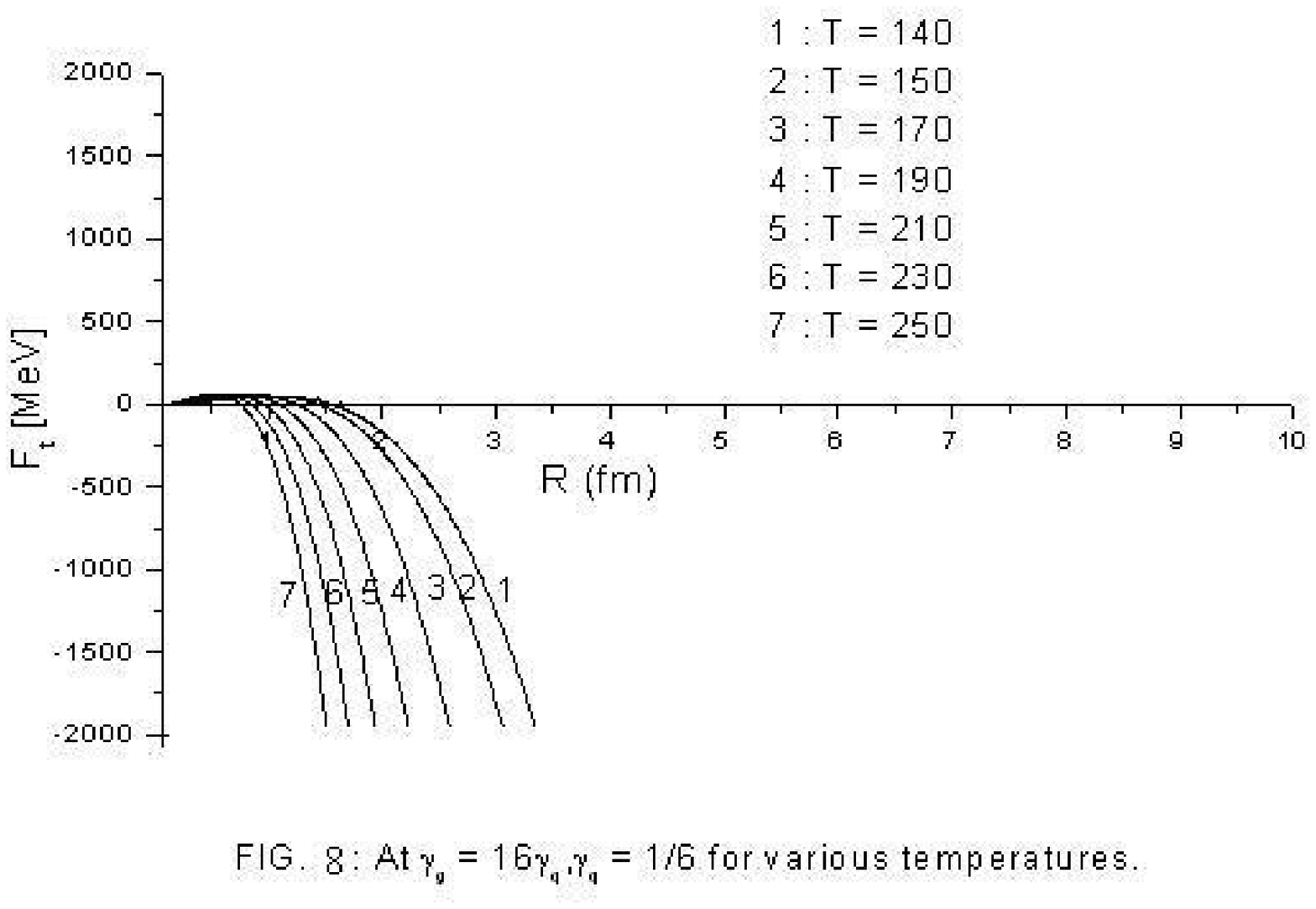,height=2.5in,width=3.5in}
\label{fig8}
\caption{The variation of total Free-energy $F_{t}$ of the QGP droplets in a pionic medium at different temperatures for the flow-parameters $\gamma_{g} = 16\gamma_{q}$ and $\gamma_{q} = 1/6$ . }
\end{center}
\end{figure}

In the present approach the bag energy is replaced by the interface energy (6) and the individual Free-energy contributions are 
shown in Figure $1$ for a particular temperature viz. $T = 152$ MeV for $\gamma_{g} = 10\gamma_{q}$, $\gamma_{q} = 1/6$. The behaviour of the total Free-Energy of the droplets with increasing droplet size for various temperatures in the range $120 ~  MeV < T < 250 ~ MeV$ for the various sets of flow-parameters  $\gamma_{q} \leq \gamma_{g} \leq 12\gamma_{q}$ with $\gamma_{q} = 1/6$ ( Peshier -Value ) are illustrated by Figure $2$ to Figure $8$.

It can be seen that the QGP-droplet-Hadron Free-energy goes on  increasing without any stable droplet forming for a choice of the flow-parameters $\gamma_{q} \leq \gamma_{g} \leq 5\gamma_{q}$,with $\gamma_{q}$ fixed at the value $1/6$ as is evident from the graphs Figure $2$ to  Figure $3$. Large stable QGP droplets of $R > 6$ fm start appearing for the value of $\gamma_{g} = 6\gamma_{q}$ at $T > 140$ MeV [Figure 4]. Stable  QGP droplets with smaller radii less than $6$ fm start appearing for a choice of $\gamma_{g} > 6\gamma_{q}$, albeit with much lower barrier heights indicating  that the droplets are highly unstable and the QGP-hadron phase transition occurs at lower temperatures of $T \sim 170 ~ MeV$ [Figure 5 to Figure 7]. At $ \gamma_{g} > 12\gamma_{q}$ [Figure 8] the droplets become highly unstable with the barrier height almost vanishing,  so that the system spontaneously passes into a QGP phase without the intermediate state of QGP droplet formation at much lower temperatures of $T < 100 ~ MeV$. The crucial role played by the hydrodynamical flow-parameters indicates both their need and primacy in adapting a statistical model meant for a cold system of electrons or nucleons to an essentially hot plasma system of QGP. Also the smooth cut at the phase boundary is indicative of a first -order phase transition as suggested by earlier authors using other models [3,4]. In short the model gives a 
simple and robust mechanism for the transition from the hadronic phase to the QGP phase with a minimal phenomenological input in terms of the hydrodynamical flow-parameters and the current quark masses. But as to which of the scenario occurs in actuality , only experiments can 
tell. The occurance of droplets with relative stability with a radius of $\sim$  $6$ fm at $\gamma_{q} \sim 1/6$ and $\gamma_{g} \sim 1$ with transition temperature $>$ $150$ MeV makes this choice of the flow parameter values most appropriate and in agreement with Lattice Gauge expectations.

\n {\bf Acknowledgement:} 

One of the authors, Agam K. Jha, would like to express his gratitude to Council of Scientific and Industrial Research (CSIR), India for financial assistance.

\n {\bf References :}
\begin{enumerate}
\item{F. Karsch, E. Laermann, A. Peikert, Ch. Schmidt and S. Stickan, Nucl. Phys. B (Proc. Suppl.) 94, 411 (2001)}.
\item{T. Renk, R. Schneider, and W. Weise, Phys. Rev. C 66, 014902 (2002)}.
\item{R. Balian and C. Block, Ann. Phys. (N.Y.) 64, 401 (1970)}.
\item{G. Neergaard and J. Madsen, Phys. Rev. D 60, 054011 (1999)}.
\item{W. Cassing, W. Ehehalt and C.M. Ko, Phys. Lett. B 363, 35 (1995)}.
\item{G. Q. Li, C.M. Ko, G. E. Brown and H. Sorge, Nucl. Phys. A 611, 539 (1996)}.
\item{J. Sollfrank, P. Huovinen, M. Kataja, P. V. Ruuskanen, M. Prakash, and R. Venugopalan, Phys. Rev. C 55, 392 (1997)}.
\item{C. M. Hung and E. Shuryak, Phys. Rev. C 57, 1891 (1998); E. Shuryak, hep-ph 0312227}.
\item{F. Becattini, J. Cleymans, A. Keranen, E. Suhonen and K. Redlich, hep-ph/0002267, Phys. Rev. C 64, 024901 (2001)}.
\item{G. D. Yen and M.I. Gorenstein, Phys. Rev. C 59, 2788 (1999); A. Peshier, B. Kampfer, O. P. Pavlenko and G. Soff, Phys. Lett. B 337, 235 (1994)}. 
\item{R. Ramanathan, Y. K. Mathur, K. K. Gupta, Proc. IVth Int. Conf. on QGP, Jaipur (2001)}.
\item{R. Ramanathan, Y. K. Mathur, K. K. Gupta and Agam K. Jha, hep-ph-0402272}.
\item{E. Fermi, Zeit F. Physik 48, 73 (1928); L. H. Thomas , Proc. camb. Phil. Soc. 23, 542 (1927)}.
\item{H. A. Bethe, Rev. Mod. Phys. 9, 69 (1937)}.
 \end{enumerate}

\end{document}